\documentclass[a4paper]{article}
\usepackage{geometry}
\usepackage{amsmath,graphicx,color}  
\usepackage{amssymb}
\usepackage{subcaption}
\usepackage{array}
\usepackage{bm}
\usepackage{calrsfs}
\usepackage{multicol}
\usepackage{tcolorbox}
\usepackage{cancel}
\usepackage{textcomp}

\newcommand{\eqnref}[1]  {equation~(\ref{#1})}

\newcommand{\secref}[1]  {section~\ref{#1}}

\DeclareMathOperator{\E}{\mathbb{E}}

\definecolor{darkgreen}{rgb}{0.0, 0.6, 0.0}

\title{Kalman filters as the steady-state solution of gradient descent on variational free energy}
\author{Manuel Baltieri$^{1, *}$, Takuya Isomura$^2$\\
\mbox{}\\
$^1$ Araya Inc., Tokyo, Japan\\ \\
$^2$ Brain Intelligence Theory Unit, \\RIKEN Centre for Brain Science, Saitama, Japan \\ \\
$^*$ Corresponding author: manuel\_baltieri@araya.org} 

\date{}

\begin{document}

\maketitle

\begin{abstract}
    The Kalman filter is an algorithm for the estimation of hidden variables in dynamical systems under linear Gauss-Markov assumptions with widespread applications across different fields. Recently, its Bayesian interpretation has received a growing amount of attention especially in neuroscience, robotics and machine learning. In neuroscience, in particular, models of perception and control under the banners of predictive coding, optimal feedback control, active inference and more generally the so-called Bayesian brain hypothesis, have all heavily relied on ideas behind the Kalman filter. Active inference, an algorithmic theory based on the free energy principle, specifically builds on approximate Bayesian inference methods proposing a variational account of neural computation and behaviour in terms of gradients of variational free energy. Using this ambitious framework, several works have discussed different possible relations between free energy minimisation and standard Kalman filters. With a few exceptions, however, such relations point at a mere qualitative resemblance or are built on a set of very diverse comparisons based on purported differences between free energy minimisation and Kalman filtering. In this work, we present a straightforward derivation of Kalman filters consistent with active inference via a variational treatment of free energy minimisation in terms of gradient descent. The approach considered here offers a more direct link between models of neural dynamics as gradient descent and standard accounts of perception and decision making based on probabilistic inference, further bridging the gap between hypotheses about neural implementation and computational principles in brain and behavioural sciences.
\end{abstract}

\section{Introduction}
The Kalman filter is a method for the solution of estimation problems in dynamical models and arguably one of the most significant advances in estimation and filtering theory of the last century. Estimation, or inference, refers to the task of evaluating some latent, or hidden, variables given the availability of some other measurable quantities, usually named observations. Estimation problems on dynamical (i.e., time-dependent) quantities in time series models are usually classified as either filtering, smoothing or prediction (forecasting). In filtering, estimation is defined for some hidden variable $s$ at time $t$, $s_t$, given a (time) series of observations $y$ at previous and present time steps $0 : t$, i.e., $y_{0 : t}$. Smoothing problems correspond to the estimation of a hidden variable $s_t$ given past, present and future observations $y_{0 : T}$ up to some (future) time $T$ such that $0 < t < T$. Forecasting, or prediction, on the other hand entails inference on a future hidden variable $s_t$ given only previous measurements $y_{0 : t - k}$, with $k > 0$.

The Kalman filter provides a standard tool to approach filtering problems under a well-defined set of assumptions including: Gaussian white noise, known inputs and parameters, linear dynamics and observation laws and quadratic cost functions. Over the years, the Kalman filter has been (re)derived using a number of different methods, e.g., orthogonal projections, innovations approach, partial differential equations, maximum likelihood and a-posterior estimates, recursive Bayesian inference, variational methods/Lagrange multipliers, hidden Markov models (for some reviews see for instance \cite{jazwinski1970stochastic, chen2003bayesian}). Recently, following the use of Bayesian principles for models in brain and behavioural studies, and the very emergence of the so-called ``Bayesian brain'' hypothesis \cite{rao1999predictive, knill2004bayesian, doya2007bayesian, clark2013whatever}, Kalman filters and a more general class of methods, i.e., Bayesian filters, have been linked to different emerging frameworks in neuroscience. These include, among others, predictive coding \cite{rao1999optimal, rao1999predictive}, optimal feedback control \cite{todorov2002optimal, todorov2005stochastic} and active inference \cite{Friston2008a, friston2017graphical}. Often, connections are proposed at a functional level, but in some cases we can also find more direct, concrete hypotheses about their neural implementations \cite{deneve2007optimal, wilson2009neural}, see in particular \cite{friedrich2021neural} for a recent review (and a new proposal).

Active inference is an ambitious framework originally developed in the natural sciences that is claimed to provide a mathematical treatment of biological, cognitive and even socio-cultural processes under the umbrella of \emph{Bayesian mechanics}, i.e., ``laws of motion'' describing spatio-temporal dynamics of parametrised Bayesian beliefs modelled in brains, biological organisms and other self-sustaining non-living systems at higher and lower scales \cite{Friston2008a, Friston2010nature, friston2019free, parr2020markov, veissiere2020thinking, baltieri2020predictions, kim2021bayesian, da2021bayesian}. In analogy with the principle of stationary action in physics, from which Newtonian, Lagrangian and Hamiltonian dynamics can be obtained under different sets of assumptions, active inference proponents claim that this framework is an attempt to derive Bayesian mechanics from a principle of (variational) free energy minimisation. Bayesian mechanics can thus essentially be seen as a set of dynamical Bayesian inference updates corresponding to a gradient descent/flow on parametrised statistical manifolds (cf. similar work on Wesserstein spaces \cite{chaudhari2018stochastic, trillos2020bayesian} building on the JKO equation \cite{jordan1998variational}).

In this light, we will then look at how such a proposal relates to established methods in Bayesian estimation, in particular whether such methods can be derived as solutions to a set of variational problems formulated with different sets of assumptions under the same Bayesian principle. This follows works such as \cite{baltieri2018probabilistic}, approximating active inference on trajectories for multiple embedding orders to obtain PID control, and builds on established results that more generally link variational treatments of Bayesian inference to classical estimation and control (via the inference/control duality) \cite{eyink2000variational, mitter2003variational, daunizeau2009variational}.

While different connections between Kalman, and more in general Bayesian, filters and active inference have been proposed in the literature, see \secref{sec:relatedWork}, only a few have managed to show a clear and formal relation \cite{de2017factor, van2021application}, specifically using factor graphs. The goal of this work is to further elucidate this connection under a different light, with a variational (Gaussian) treatment of Bayesian filters, presented in \secref{sec:BayesianFilter}, followed by a generative model and related free energy function for a Kalman filter in \secref{sec:VFEKalmanFilter}. In \secref{sec:minimisationFreeEnergy} we then derive a standard gradient descent/flow of free energy minimisation as specified by the Bayesian mechanics formulation of active inference \cite{Friston2008a, Friston2010nature, friston2019free, parr2020markov, veissiere2020thinking, kim2021bayesian, da2021bayesian}. As we shall see, Kalman filters will emerge as the steady state solution of a gradient descent on variational free energy under the right generative model. This derivation will allow us to draw more direct connections to gradient-based frameworks in different fields, see concluding remarks in \secref{sec:conclusion}, highlighting then their role in neuroscience as a possible connection between models of neural dynamics relying on gradient descent schemes and computational principles describing cognitive functions as Bayesian belief updates.



\section{Related work}
\label{sec:relatedWork}
Several works in the literature propose different --- and at times contradicting --- relations between active inference and Kalman or Kalman-Bucy filters (the discrete observations continuous dynamics version of Kalman filters). For instance, some works suggest that Kalman filters are equivalent to predictive coding (itself claimed to be a special case of active inference), at least in some cases and under some implicit assumptions \cite{bastos2012canonical, moran2013free, corcoran2020allostatic, barron2020prediction}. It is then often claimed that Kalman or Kalman-Bucy filters are superseded, generalised or entailed by the free energy principle and/or active inference \cite{Friston2008a, Friston2008c, adams2012smooth, friston2013free, adams2013psychosis, adams2014bayesian, kneissler2015simultaneous}, but a formal derivation is usually missing. At times, Kalman(-Bucy) and the more general Bayesian filters are said to ``resemble'', be ``formally similar to'' or ``consistent with'', ``have the same form as'' or ``take the form of'' equations in active inference \cite{kiebel2011free, friston2012dopamine, friston2012value, friston2012free, friston2012freeValue, friston2012embodied, friston2012perception, friston2012perceptions, adams2012smooth, brown2013active, perrinet2014active, friston2014cognitive, pio2016active, friston2019free, parr2020markov, meera2020free}. Other works suggest that active inference proposes a form of Bayesian filtering and in some cases generalises it \cite{Friston2010genfilt, friston2015active, friston2015knowing, friston2017graphical, parr2018active, isomura2019bayesian, palacios2020markov, da2021bayesian}\footnote{Notice that in these accounts, Kalman filters are not mentioned directly as in other references, but their connection can be inferred from the very definition of Bayesian filters.}. Related arguments are sometimes used to claim that generative models can be defined to support both active inference and Kalman filtering \cite{parr2019neuronal}, although the exact definition of such generative models is not provided. In some cases, it is also stated that using variational free energy minimisation one can derive Kalman filters \cite{kuchling2020morphogenesis}, but to the best of our knowledge, no explicit proof of a gradient-based approach consistent with active inference and the free energy principle has been provided. On the other hand, Kalman filters have been formally described via a factor graphs description of active inference algorithms \cite{de2017factor, van2021application} relating to exact treatments of message passing \cite{roweis1999unifying}.

Other works then argue that Kalman filters and variational free energy derivations in active inference are distinct and in some cases can be compared to each other \cite{Friston2008a, Friston2008c, balaji2011bayesian, li2011generalised, kneissler2015simultaneous, lanillos2018adaptive, meera2020free, baltieri2020kalman, baioumy2021fault, bos2021free, anil2021dynamic}, often claiming that active inference formulations outperform Kalman filters \cite{Friston2008a, Friston2008c, balaji2011bayesian, li2011generalised, kneissler2015simultaneous, meera2020free, bos2021free, anil2021dynamic}. In \cite{millidge2020relationship, millidge2021predictive} we then also find claims that approximations of Kalman filters derived from active inference are supposedly more biological plausible than their counterpart in standard filters. Finally, due to the connections between active inference and control as inference, Kalman filters have also been contrasted to active inference in \cite{imohiosen2020active}, even if their exact relation remains unstated.

In this literature we find that only but a few works attempted to provide explicit formal connections via gradient descent on variational free energy \cite{baltieri2020kalman, meera2020free, millidge2020relationship, millidge2021predictive}, or via message passing on factor graphs \cite{de2017factor, van2021application}. Even fewer then have managed to correctly capture the actual equivalence \cite{de2017factor, van2021application}, none of which using gradient-based algorithms. To address this series of inconsistencies, and to expand on previous results, in what follows we derive state estimation for continuous-state distributions under active inference, with a focus on filtering algorithms. For a closely related variational treatment, relying on algebraic manipulations over gradients and with applications also to smoothing, regression and other inference problems, see also \cite{ostwald2014tutorial}.

\section{A variational (Gaussian) treatment of Bayesian filters}
\label{sec:BayesianFilter}
The standard formulation of Bayesian filtering problems for linear Gauss-Markov models can be described in terms of recursive Bayesian estimates \cite{chen2003bayesian, haykin2009neural, sarkka2013bayesian} over hidden states $s_t$ with observations $y_{0 : t}$ for $t= 0 \dots T$:
\begin{align}
    p(s_{t} | y_{0 : t}) & = \frac{p(y_{0 : t}, s_{t})}{p(y_{0 : t})} \nonumber \\
    & = \frac{p(y_{t}, y_{0 : t-1}, s_{t})}{p(y_{t}, y_{0 : t-1})} \nonumber \\
    & = \frac{p(y_{t} | s_{t}, y_{0 : t-1}) p(s_{t} | y_{0 : t-1}) p(y_{0 : t-1})}{p(y_{t} | y_{0 : t-1}) p(y_{0 : t-1})} \nonumber \\
    & = \frac{p(y_{t} | s_{t}) p(s_{t} | y_{0 : t-1})}{p(y_{t} | y_{0 : t-1})} 
    \label{eqn:generativeModelProbabilistic}
\end{align}
where in the last line we used the fact that observations at time $t$, $y_t$, are conditionally independent of their past given $s_t$.
%
%
From this, we can then derive a variational treatment establishing a bound for the surprisal (i.e., the negative log-model evidence), $- \ln p(y_{t} | y_{0 : t-1})$, introducing a variational density $q(s_t)$ 
\begin{align}
    - \ln p(y_{t} | y_{0 : t-1}) & = \ln p(s_t | y_{t}, y_{0 : t-1}) - \ln p(y_{t}, s_t | y_{0 : t-1}) \nonumber \\
    & = \ln p(s_t | y_{t}, y_{0 : t-1}) - \ln p(y_{t}, s_t | y_{0 : t-1}) + \ln q(s_t) - \ln q(s_t) \nonumber \\
    & = \ln \frac{q(s_t)}{p(y_{t}, s_t | y_{0 : t-1})} - \ln \frac{q(s_t)}{p(s_t | y_{t}, y_{0 : t-1})} \nonumber \\
    & = \int q(s_t) \ln \frac{q(s_t)}{p(y_{t}, s_t | y_{0 : t-1})} \, d s_t - \int q(s_t) \ln \frac{q(s_t)}{p(s_t | y_{t}, y_{0 : t-1})} \, d s_t
    \label{eqn:surprisal}
\end{align}
and defining the variational free energy as
\begin{align}
    F & \equiv \int q(s_t) \ln \frac{q(s_t)}{p(y_{t}, s_t | y_{0 : t-1})} \, d s_t \nonumber \\
    & = \E_{q(s_t)}[-\ln p(y_{t}, s_t | y_{0 : t-1}) + \ln q(s_t)]
\end{align}
where $\E_{q(s_t)}[\bullet]$ denotes the expectation of $\bullet$ over $q(s_t)$. This gives an upper bound of the surprisal owing to the non-negativity of the Kullback-Leibler divergence in the second term of the last line of \eqnref{eqn:surprisal}. Under a variational Gaussian assumption \cite{opper2009variational}, the variational density is defined as
\begin{align}
    q(s_t) = \mathcal{N}(s_t; \mu_{s_t}^+, \Sigma_{s_t}^+) = (2 \pi)^{-N/2} \, |\, \Sigma_{s_t}^+ \, |^{-1/2} e^{-\frac{1}{2}(s_t - \mu_{s_t}^+)^T \left( \Sigma_{s_t}^+ \right)^{-1} (s_t - \mu_{s_t}^+)}
\end{align}
The ``$^+$-notation'' is introduced here to compare this derivation with standard Kalman filter treatments differentiating between parameters that take into account information about observations $y_t$ at time $t$ (corrections, or a-posteriori estimates of $s_t$) and parameters that don't (predictions, or a-priori estimates of $s_t$ for which we will later adopt a ``$^-$-notation''). Using textbook results for the differential entropy of a multivariate Gaussian distribution, the variational free energy reduces to
\begin{align}
    F 
    = \E_{q(s_t)}[- \ln p(y_{t}, s_t | y_{0 : t-1})] - \frac{N}{2} \ln (2 \pi e) - \frac{1}{2} \ln |\, \Sigma_{s_t}^+ \, |
    \label{eqn:variationalGaussian}
\end{align}
Following \cite{Friston2008a}, we then apply a Laplace approximation to the joint $p(y_{t}, y_{0 : t-1}, s_t)$\footnote{Notice that with the variational treatment applied in \cite{Friston2008a}, the posterior is effectively of the form: $e^{\E_{q(s_t)}[- \ln p(y_{t}, y_{0 : t-1}, s_t)]}$, which entails the equivalence of the variational Gaussian approximation applied to $q(s_t)$ and the Laplace method applied to $p(y_{t}, y_{0 : t-1}, s_t)$ \cite{opper2009variational}.} \cite{mackay2003information}. In particular, this reduces to a Taylor expansion up to second order of the log-joint distribution at the maximum a posteriori (mode) of the posterior, $\hat{s}_t$
\begin{align}
    \ln p(y_{t}, s_t | y_{0 : t-1}) = \ln p(y_{t}, s_t | y_{0 : t-1}) \Big \rvert_{s_t = \hat{s}_t} - \frac{1}{2} (s_t - \hat{s}_t)^T S_t^{-1} (s_t - \hat{s}_t) + \dots
\end{align}
with the first order term not appearing (i.e., $= 0$) at the expansion point, the mode, and with the matrix $S_t^{-1}$ equal to the Hessian of $\ln p(y_{t}, y_{0 : t-1}, s_t)$) evaluated at the mode $s_t = \hat{s}_t$
\begin{align}
    & \nabla_{s_t} \ln p(y_{t}, s_t | y_{0 : t-1}) \Big \rvert_{s_t = \hat{s}_t} = 0 \nonumber \\
    & \nabla_{s_t} \nabla_{s_t} \ln p(y_{t}, s_t | y_{0 : t-1}) \Big \rvert_{s_t = \hat{s}_t} = - S_t^{-1} \nonumber \\
    & \Rightarrow S_t^{-1} = - \nabla_{s_t} \nabla_{s_t} \ln p(y_{t}, s_t | y_{0 : t-1}) \Big \rvert_{s_t = \hat{s}_t}
\end{align}
At this stage, the free energy can be rewritten, omitting higher order terms, as
\begin{align}
    F \approx & -\ln p(y_{t}, s_t | y_{0 : t-1}) \Big \rvert_{s_t = \hat{s}_t} + \frac{1}{2} \E_{q(s_t)} \left[(s_t - \hat{s}_t)^T S_t^{-1} (s_t - \hat{s}_t) \right] - \frac{N}{2} \ln (2 \pi e) - \frac{1}{2} \ln |\, \Sigma_{s_t}^+ \, |
    \label{eqn:LaplaceFreeEnergyBeforeAssumption}
\end{align}
%
Following standard treatments, \cite{Friston2008a, opper2009variational} we will simplify the above free energy $F$ by considering the case where the variational density $q(s_t)$ is centred at the mode of the posterior $\hat{s}_t$, i.e., $\mu_{s_t}^+ = \hat{s}_t$, and where $\Sigma_{s_t}^+ = S_t$, i.e., the covariance of the variational density $\Sigma_{s_t}^+$ is equal to the covariance of the generative density $S_t$ at each expansion point (i.e., the mode) of the Laplace method. 
%
This then implies that, using the trace trick,
\footnote{Notice that the equality holds strictly only for the case where both $q(s_t)$ and $p(y_{t}, s_t | y_{0 : t-1})$ are Gaussian.}
\begin{align}
    F \approx & -\ln p(y_{t}, s_t | y_{0 : t-1}) \Big \rvert_{s_t = \hat{s}_t} - \frac{N}{2} \ln (2 \pi) - \frac{1}{2} \ln |\, \Sigma_{s_t}^+ \, | 
    \label{eqn:LaplaceFreeEnergy}
\end{align}

\section{Variational free energy for a Kalman filter}
\label{sec:VFEKalmanFilter}
After writing down the variational free energy (under Gaussian assumptions) for a generic Bayesian filtering problem, we now specify the components of a linear generative model, $p(y_{t}, s_t | y_{0 : t-1}) = p(y_{t} | s_t) p(s_t | y_{0 : t-1})$ as follows:
\begin{align}
    p(y_{t} | s_{t}) & = \mathcal{N} (C \mu_{s_t}^+, \Sigma_z) \nonumber \\
    p(s_{t} | y_{0 : t-1}) & = \mathcal{N} (\mu_{s_t}^{-}, \Sigma_{s_{t}}^{-})
    \label{eqn:generativeModelBrokenDown}
\end{align}
using the $^+-$ and $^--$ notations described previously. Here, the a-priori estimates are given as functions of the a-posteriori estimates at the last time step
\begin{align}
    \mu_{s_t}^{-} & \equiv A \mu_{s_{t - 1}}^+ \nonumber \\
    \Sigma_{s_{t}}^{-} & \equiv A \Sigma_{s_{t - 1}}^+ A^T + \Sigma_w
    \label{eqn:prior}
\end{align}
Notice that this can be seen, equivalently, as a linear state-space model under Gauss-Markov assumptions
\begin{align}
    s_0 & = w_0 \nonumber \\
    s_{t} & = A s_{t - 1} + w_t, \quad \text{for $t>0$} \nonumber \\
    y_t & = C s_t + z_t
    \label{eqn:generativeModel}
\end{align}
with Gaussian noise $w_t \sim \mathcal{N} (0, \Sigma_w), z_t \sim \mathcal{N} (0, \Sigma_z)$. In probabilistic form this is then
\begin{align}
    p(s_1 | s_0) & = \mathcal{N} (0, \Sigma_w) \nonumber \\
    p(s_{t} | s_{t - 1}) & = \mathcal{N} (A s_{t - 1}, \Sigma_w) \nonumber \\
    p(y_{t} | s_{t}) & = \mathcal{N} (C s_t, \Sigma_z)
    \label{eqn:generativeModelProbs}
\end{align}
with equations that can be combined to form, using the Chapman-Kolmogorov equation \cite{sarkka2013bayesian} for standard Gaussian distributions (i.e., a convolution of Gaussians), the following
\begin{align}
    p(s_{t} | y_{0 : t-1}) & = \int p(s_{t} | s_{t - 1}) p(s_{t - 1} | y_{0 : t-1}) \, ds_{t - 1} \nonumber \\
    & = \int \mathcal{N} (A s_{t - 1}, \Sigma_w) \mathcal{N} (\mu_{s_{t - 1}}^+, \Sigma_{s_{t - 1}}^+) \, d s_{t - 1} \nonumber \\
    & = \mathcal{N} (A \mu_{s_{t - 1}}^+, A \Sigma_{s_{t - 1}}^+ A^T + \Sigma_w) \nonumber \\
    & = \mathcal{N} (\mu_{s_t}^{-}, \Sigma_{s_{t}}^{-})
\end{align}
as seen in \eqnref{eqn:generativeModelBrokenDown}. 

The generative model described in \eqnref{eqn:generativeModelBrokenDown} can be plugged into \eqnref{eqn:LaplaceFreeEnergy}, with $\hat{s}_t = \mu_{s_t}^+$, thus obtaining the variational free energy\footnote{Notice that, unlike in \eqnref{eqn:LaplaceFreeEnergyBeforeAssumption} and \eqnref{eqn:LaplaceFreeEnergy}, here we use an equality sign. Having by now assumed that both $q(s_t)$ and $p(y_{t}, s_t | y_{0 : t-1})$ are Gaussian, there are in fact no higher order terms to be considered in the Laplace method.}
\begin{align}
    F = & \frac{1}{2} \left[ \left( y_t - C \mu_{s_t}^+ \right)^T  \Sigma_z^{-1} \left( y_t - C \mu_{s_t}^+ \right) + \left( \mu_{s_t}^+ - \mu_{s_t}^{-} \right)^T  \left( \Sigma_{s_t}^{-} \right)^{-1} \left( \mu_{s_t}^+ - \mu_{s_t}^{-} \right) \right] + \frac{1}{2} \ln |\, \Sigma_{z_t} \, | + \frac{1}{2} \ln |\, \Sigma_{w_t} \, | \nonumber \\
    & - \frac{N}{2} \ln (2 \pi) - \frac{1}{2} \ln |\, \Sigma_{s_t}^+ \,|
    \label{eqn:freeEnergyLinearDynamicInstantaneous}
\end{align}

\section{Free energy minimisation with respect to $(\mu_{s_t}^+, \Sigma_{s_t}^+)$}
\label{sec:minimisationFreeEnergy}
A Kalman filter can be derived from the minimisation of \eqnref{eqn:freeEnergyLinearDynamicInstantaneous} with respect to $(\mu_{s_t}^+, \Sigma_{s_t}^+)$ after providing the following gradients of free energy:

\begin{align}
    \nabla_{\mu_{s_t}^+} F & = - C^T \Sigma_z^{-1} \left( y_t - C \mu_{s_t}^+ \right) + \left( \Sigma_{s_t}^{-} \right)^{-1} \left (\mu_{s_t}^+ - \mu_{s_t}^{-} \right) \nonumber \\
    \nabla_{\Sigma_{s_t}^+} F & = \nabla_{\Sigma_{s_t}^+} \left[ - \ln p(y_{t}, s_t | y_{0 : t-1}) \Big \rvert_{s_t = \mu_{s_t}^+} \right] - \frac{1}{2} \left( \Sigma_{s_t}^+ \right)^{-1} \nonumber \\ 
    & = \frac{1}{2} \nabla_{\mu_{s_t}^+} \nabla_{\mu_{s_t}^+} \left[ - \ln p(y_{t}, s_t | y_{0 : t-1}) \Big \rvert_{s_t = \mu_{s_t}^+} \right] - \frac{1}{2} \left( \Sigma_{s_t}^+ \right)^{-1} \nonumber \\
    & = \frac{1}{2} \left( C^T \Sigma_z^{-1} C + \left( \Sigma_{s_t}^{-} \right)^{-1} \right) - \frac{1}{2} \left( \Sigma_{s_t}^+ \right)^{-1}
    \label{eqn:gradients}
\end{align}
Notice that in the second equation, we applied Price's and Bonnet's theorems \cite{opper2009variational, khan2021bayesian} to compute (first order) derivatives with respect to $\Sigma_{s_t}^+$ using (second order) derivatives with respect to $\mu_{s_t}^+$.

\subsection{Covariance estimation}
The equation for the correction updates of the covariance parameters $\Sigma_{s_t}^+$ can be obtained starting from the steady-state solution for $\nabla_{\Sigma_{s_t}^+} F = 0$
\begin{align}
    \left( \Sigma_{s_t}^+ \right)^{-1} & = C^T \Sigma_z^{-1} C + \left( \Sigma_{s_t}^{-} \right)^{-1} \nonumber \\
    \Rightarrow \Sigma_{s_t}^+ & = \left( C^T \Sigma_z^{-1} C + \left( \Sigma_{s_t}^{-} \right)^{-1} \right)^{-1} \nonumber \\
    & = \Sigma_{s_t}^{-} - \Sigma_{s_t}^{-} C^T \left( \Sigma_z + C \Sigma_{s_t}^{-} C^T \right)^{-1} C \Sigma_{s_t}^{-} \nonumber \\
    & = (I - K_{s_t} C) \Sigma_{s_t}^{-}
    \label{eqn:covarianceCorrection}
\end{align}
where in the second line we applied the standard matrix inversion (Sherman–Morrison–Woodbury) lemma and in the last line we used the familiar definition of the \emph{Kalman gain} \cite{simon2006optimal}
\begin{align}
    K_{s_t} & = \Sigma_{s_t}^{-} C^T \left( \Sigma_z + C \Sigma_{s_t}^{-} C^T \right)^{-1}
    \label{eqn:KalmanGain}
\end{align}
The definition of $\Sigma_{s_t}^{-}$ in \eqnref{eqn:prior} reflects then the prediction step
\begin{align}
    \Sigma_{s_t}^{-} = A \Sigma_{s_{t - 1}}^{+} A^T + \Sigma_w
    \label{eqn:covariancePrediction}
\end{align}
and can be substituted in \eqnref{eqn:covarianceCorrection} to obtain a standard one-step update equation \cite{simon2006optimal}.

\subsection{Mean estimation}
For expectation parameters $\mu_{s_t}^+$, once again at steady-state ($\nabla_{\mu_{s_t}^+} F = 0$), we obtain after a few manipulations the following variational updates
\begin{align}
    0 = & - C^T \Sigma_z^{-1} \left( y_t - C \mu_{s_t}^+ \right) + \left( \Sigma_{s_t}^{-} \right)^{-1} \left (\mu_{s_t}^+ - \mu_{s_{t}}^- \right) \nonumber \\
    \left( C^T \Sigma_z^{-1} C + \left( \Sigma_{s_t}^{-} \right)^{-1} \right) \mu_{s_t}^+ = & C^T \Sigma_z^{-1} y_t + \left( \Sigma_{s_t}^{-} \right)^{-1} \mu_{s_{t}}^- \nonumber \\
    \left( (I - K_{s_t} C) \Sigma_{s_t}^{-} \right)^{-1} \mu_{s_t}^+ = & C^T \Sigma_z^{-1} y_t + \left( \Sigma_{s_t}^{-} \right)^{-1} \mu_{s_{t}}^- \nonumber \\
    \mu_{s_t}^+ = & \mu_{s_{t}}^- + (I - K_{s_t} C) \Sigma_{s_t}^{-} C^T \Sigma_z^{-1} \left( y_t - C \mu_{s_{t}}^- \right) \nonumber \\
    = & \mu_{s_{t}}^- + K_{s_t} ( y_t - C \mu_{s_{t}}^- ) 
    \label{eqn:meanCorrection}
\end{align}
where in the last line we used the equivalent form of the Kalman gain in \eqnref{eqn:KalmanGain}
\begin{align}
    K_{s_t}\left( \Sigma_z + C \Sigma_{s_t}^{-} C^T \right) & = \Sigma_{s_t}^{-} C^T \nonumber \\
    K_{s_t} \left( I + C \Sigma_{s_t}^{-} C^T \Sigma_z^{-1} \right) & = \Sigma_{s_t}^{-} C^T \Sigma_z^{-1} \nonumber \\
    K_{s_t} & = \Sigma_{s_t}^{-} C^T \Sigma_z^{-1} - K_{s_t} C \Sigma_{s_t}^{-} C^T \Sigma_z^{-1} \nonumber \\
    & = (I - K_{s_t} C) \Sigma_{s_t}^{-} C^T \Sigma_z^{-1} \nonumber \\
    & = \Sigma_{s_t}^{+} C^T \Sigma_z^{-1}
\end{align}
The definition of $\mu_{s_t}^{-}$ in \eqnref{eqn:prior} then implements the prediction step:
\begin{align}
    \mu_{s_t}^{-} = A \mu_{s_{t - 1}}^+
\end{align}
and, similarly to the covariance updates, can be plugged in \eqnref{eqn:meanCorrection} for a one-step update rule that concludes our derivation.

\section{Concluding remarks}
\label{sec:conclusion}
Kalman filters are a cornerstone method for state estimation of dynamical systems under linear Gauss-Markov assumptions and are often nowadays formulated as a process of recursive Bayesian inference. Active inference is a framework originally developed in neuroscience with the goal of describing processes of neural computation and behavioural features of cognitive agents as approximate Bayesian inference \cite{Friston2010nature, friston2015active, friston2017active}. More recently, this framework has been explicitly formalised in terms of \emph{Bayesian mechanics}, equations of motion for parameter updates in a statistical manifold. In practice, these equations correspond to a gradient descent/flow in variational free energy schemes, usually under Gaussian assumptions \cite{friston2019free, kim2021bayesian, da2021bayesian}.

In this note we shed light on a connection between Kalman filters and active inference, specifically its formulation in terms of gradient descent/flow. In particular, we showed that Kalman filters are the steady-state solution of a gradient descent scheme on the parameters (means and covariance matrix) of a linear generative model under Gaussian assumptions (Laplace approximation/variational Gauss \cite{opper2009variational}). This complements existing algebraic formulations \cite{ostwald2014tutorial}, including the ones based on message passing algorithms \cite{de2017factor, van2021application}, and clarifies the role of Kalman filters as gradients of variational free energy, providing a precise connection between standard filtering methods and gradients in Bayesian belief space.

Importantly, this derivation should not be confused with the idea of ``steady-state (Kalman) filtering'' \cite{simon2006optimal}: our derivation shows a generic Kalman filter as the steady-state solution of free energy minimisation, while steady-state filters refer to methods where the dynamics of the filters themselves reach steady-state, i.e., the covariance $\Sigma_{s_t}^+ $ becomes stationary.

With the assumption of \emph{linear} generative models, it would then be natural also to question the necessity of a variational treatment in the first place, given the usual interpretation of variational Bayes as a method to provide an approximate bound for the model evidence of an analytically or numerically intractable problem \cite{jordan1999introduction, beal2003variational, bishop2006pattern, Friston2008a}. Here however, we take a different perspective and see the current work as inspired by treatments such as \cite{jordan1998variational}, highlighting the tight connections between geometric (gradient descent on statistical manifolds), variational (bounds in terms of free energy or ELBO) and Bayesian inference interpretations of filtering. This approach can be seen for example in complementary works for Kalman filters in continuous-time (both dynamics and observations) such as \cite{halder2017gradient, halder2018gradient}.

In this light, the present variational Bayesian derivation of the Kalman filter shows as a way to 1) preserve its classical connections to Bayesian inference for the exact (i.e., linear) case \cite{ho1964bayesian}, 2) understand the Kalman filter as the optimal linear estimator for non-linear systems \cite{chen2003bayesian, sarkka2013bayesian} in terms of a fixed-form (variational) Gaussian approximation of arbitrary distributions \cite{opper2009variational, honkela2010approximate} and 3) link the filter to more recent (information) geometric treatments of probabilistic inference \cite{li2017bayesian, ollivier2019extended}.

In addition, our gradient-based approach preserves, more directly, (potential) connections that have been proposed between active inference and frameworks in, e.g., physics \cite{friston2019free, baltieri2020predictions, kim2021bayesian}, chemistry \cite{parr2021message}, and brain science, specifically for neural dynamics \cite{friston2017active, whittington2017approximation, whittington2019theories, isomura2020canonical, da2021neural}. Finally, this clarifies a series of claims about the relation between Kalman filters and active inference \cite{Friston2008a, Friston2008c, Friston2010genfilt, balaji2011bayesian, li2011generalised, kiebel2011free, bastos2012canonical, friston2012free, friston2012perception, friston2012perceptions, friston2012embodied, friston2012dopamine, friston2012value, friston2012freeValue, adams2012smooth, adams2013psychosis, friston2013free, moran2013free, brown2013active, adams2014bayesian, kneissler2015simultaneous, perrinet2014active, friston2014cognitive, friston2015active, pio2016active, friston2015knowing, de2017factor, friston2017graphical, lanillos2018adaptive, parr2018active, isomura2019bayesian, friston2019free, parr2019neuronal, parr2020markov, meera2020free, corcoran2020allostatic, barron2020prediction, palacios2020markov, kuchling2020morphogenesis, baltieri2020kalman, imohiosen2020active, millidge2020relationship, millidge2021predictive, baioumy2021fault, van2021application, da2021bayesian, bos2021free, anil2021dynamic}, resolving puzzles derived from ambiguous and inconsistent treatments found in the literature.

\section{Acknowledgments}
The authors would like to thank Lancelot Da Costa for constructive feedback on a previous draft of the manuscript.

\footnotesize
\bibliographystyle{plain}
\bibliography{KFFreeEnergy} 

\end{document}